\documentclass[twocolumn,draft,showkeys,showpacs,eqsecnum,nofootinbib,aps]{revtex4}
\renewcommand{\theequation}{\arabic{equation}}
\def\beq{\begin{equation}}
\def\eeq{\end{equation}}
\def\bea{\begin{eqnarray}}
\def\eea{\end{eqnarray}}\def\nn{\nonumber}

\def\na{\nabla}
\def\pa{\partial}

\def\nn{\nonumber}

\def\ra{\rightarrow}

\begin{document}
\title{Morse theory in path space}
\author{Yong Seung Cho}
\email{yescho@ewha.ac.kr} \affiliation{National Institute for
Mathematical Sciences, 385-16 Doryong, Yuseong, Daejeon 305-340
Korea}\affiliation{Department of Mathematics, Ewha Womans
University, Seoul 120-750 Korea}
\author{Soon-Tae Hong}
\email{soonhong@ewha.ac.kr} \affiliation{Department of Science
Education and Research Institute for Basic Sciences, Ewha Womans
University, Seoul 120-750 Korea}

\date{\today}
\begin{abstract}
We consider the path space of a curved manifold on which a point
particle is introduced in a conservative physical system with
constant total energy to formulate its action functional and
geodesic equation together with breaks on the path. The second
variation of the action functional is exploited to yield the
geodesic deviation equation and to discuss the Jacobi fields on
the curved manifold. We investigate the topology of the path space
using the action functional on it and its physical meaning by
defining the gradient of the action functional, the space of
bounded flow energy solutions and the moduli space associated with
the critical points of the action functional.  We also consider
the particle motion on the $n$-sphere $S^{n}$ in the conservative
physical system to discuss explicitly the moduli space of the path
space and the corresponding homology groups.

\end{abstract}
 \maketitle

\section{Introduction}
\setcounter{equation}{0}
\renewcommand{\theequation}{\arabic{section}.\arabic{equation}}

In order to discuss the Morse
inequalities~\cite{morse34,milnor63,wald84}, the supersymmetric
quantum mechanics has been exploited by Witten~\cite{witten82}.
Based on the spectral flow of the Hessian of the symplectic
function~\cite{floer}, the Morse indices for pair of critical
points of the symplectic action functional have been also
investigated and on the Hilbert spaces the Morse
homology~\cite{schwarz} has been yielded by considering the flows
of the critical points associated with the Morse
index~\cite{majer}.

It has been noted~\cite{rho83,hong02} in the hadron phenomenology
using the chiral bag model that the quark phase spectrum is
asymmetric about zero energy to yield the nonvanishing vacuum
contribution to the baryon number.  To obtain this vacuum
contribution, the regularization has been exploited and it is
closely related~\cite{goldstone83} to the eta invariant of Atiyah
et al.~\cite{atiyah75}. The eta invariant has been later discussed
by Witten in connection with the phase factor of the path integral
in quantum field theory associated with the Jones polynomial and
knot theory~\cite{witten89}.  To relate the phase factor of the
semiclassical partition function to the eta
invariant~\cite{atiyah75}, the Jacobi fields and their eigenvalues
of the Sturm-Liouville operator associated with the particle
geodesics on a curved manifold  have been also
investigated~\cite{hong03}.

To yield geometric invariants of smooth four-manifolds, the moduli
space of self-dual connections which are critical points of the
Yang-Mills functional in SU(2) gauge
theory~\cite{donalson83,cho91} and the moduli space of solutions
of Seiberg-Witten monopole equations in U(1) gauge
theory~\cite{witten94,cho97} have been investigated.  Recently,
the Morse theoretical approach has been also applied to the
Nambu-Goto string action functional to study the geodesic surface
equation with the world sheet currents~\cite{cho07}. Constructing
the second variation of the surface spanned by closed strings, the
geodesic surface deviation equation has been discussed on the
curved manifold, and the geodesic surface deviation equation in
the orthonormal gauge has been derived to find the Jacobi field
and the conjugate strings on the geodesic surface.

In this paper we will investigate the physical changes of the
action by studying the geometry of the moduli space associated
with the critical points of the action functional and the
asymptotic boundary conditions in path space for point particles
in a conservative physical system with constant total energy,
after formulating the geodesic equation and geodesic deviation
equation together with breaks on the path. Explicitly we will
study the particle motion on the $n$-sphere $S^{n}$ in the
conservative physical system to discuss the moduli space of the
path space and the corresponding homology groups associated with
the boundary homomorphism.

In Section II, the action functional for a point particle will be
introduced in a conservative physical system with constant total
energy to investigate the geodesic equation together with breaks
on the path. By taking the second variation of the action
functional generated by point particles, the geodesic deviation
equation will be discussed in terms of the Jacobi fields on the
curved manifold. In Section III, exploiting the gradient of the
action functional, the space of bounded flow energy solutions will
be investigated together with the moduli space associated with the
critical points of the action functional and the asymptotic
boundary conditions.  The boundary homomorphism will be also
introduced to define the homology group of the path space. In
Section IV, the particle in a conservative physical system will be
considered on the $n$-sphere $S^{n}$ to discuss explicitly the
moduli space of the path space and the corresponding homology
groups.

\section{Morse index in path space}
\setcounter{equation}{0}
\renewcommand{\theequation}{\arabic{section}.\arabic{equation}}

We consider a particle of mass $m$ on a curved $n$-dimensional
manifold with metric $\eta_{ab}$ ($a,b=1,2,\cdots,n$) in a
conservative physical system with constant total energy $E=T+V$,
where $T$ and $V$ are the kinetic and potential energies,
respectively.  We then define the line element
$ds^{2}=-dt^{2}+g_{ab}dx^{a}dx^{b}$ with a dressed metric $g_{ab}$
associated with $E$ and $V$ to yield\footnote{In
Ref.~\cite{hong03}, a flat $n$-dimensional manifold with metric
$\delta_{ab}$ ($a,b=1,2,\cdots,n$), instead of $\eta_{ab}$, was
introduced in a conservative physical system with constant total
energy.} \beq
g_{ab}=\frac{m(E-2V)^{2}}{2(E-V)}\eta_{ab}.\label{metric}\eeq In
order to define an action functional on the curved manifold, let
$(M, g_{ab})$ be the complete Riemannian manifold of dimension $n$
associated with the metric $g_{ab}$. Given $g_{ab}$, we can have a
unique covariant derivative $\na_{a}$ satisfying~\cite{wald84}
$\na_{a}g_{bc}=0$,
$\na_{a}\omega^{b}=\pa_{a}\omega^{b}+\Gamma^{b}_{~ac}~\omega^{c}$
and \beq
(\na_{a}\na_{b}-\na_{b}\na_{a})\omega_{c}=R_{abc}^{~~~d}~\omega_{d}.\eeq
Let $\Omega(M;p,q)$ be the set of piecewise smooth paths $\gamma
(\tau)$ such that  $\gamma: [0,1]\rightarrow M$ and propagate from
point $p$ to $q$ in $M$. On the manifold $M$ the action functional
$S:\Omega(M;p,q)\rightarrow {\mathbf R}$ of the particle is given
by \beq S=\int_{0}^{1}d\tau~(g_{ab}v^{a}v^{b})^{1/2},
\label{action} \eeq with the proper time $\tau$ $(0\le\tau\le 1)$
and the vector field $v^{a}=(\partial/\partial \tau)^{a}\in
T_{\gamma}\Omega (M;p,q)$. Here one notes that, without loss of
generality, $V(x^{i})$ can be chosen to vanish at starting point
at $\tau=0$ and the metric $g_{ab}$ in (\ref{metric}) does not
have any singularities since its denominator is positive definite.

Let the vector field $w^{a}=(\pa/\pa \alpha)^{a}\in
T_{\gamma}\Omega (M;p,q)\subset M$ be the deviation vector which
comes from a variation $\bar{\gamma}: (-\epsilon,\epsilon)\times
[0,1] \rightarrow \Omega(M;p,q)\subset M$ such that
$\bar{\gamma}(0,\tau)=\gamma(\tau)$ and represents the
displacement to an infinitesimally nearby path. Next let $\Sigma$
denote the two-dimensional submanifold spanned by the paths
$\bar{\gamma}(\alpha)$. We now may choose $\tau$ and $\alpha$ as
coordinates of $\Sigma$ to yield the commutator relation, \beq
\pounds_{v}w^{a}=v^{b}\na_{b}w^{a}-w^{b}\na_{b}v^{a}=0.
\label{poundvw} \eeq The tangent space $T_{\gamma}\Omega(M;p,q)$
of $\Omega(M;p,q)$ at a path $\gamma(\tau)$ will be then the
vector space of all piecewise smooth vector fields $w^{a}$ along
$\gamma(\tau)$ for which \beq w^{a}(0)=w^{a}(1)=0.\label{bcwa}\eeq

Now we perform an infinitesimal variation of the paths
$\gamma(\tau)$ traced by the particle during its evolution in
order to find the geodesic equation from the least action
principle. In the stationary phase approximation where $|w^{a}|$
is infinitesimally small, we find the first variation as follows
\beq S^{(1)}=\frac{\pa S}{\pa\alpha}
=\frac{1}{c}v^{a}w_{a}|_{\tau=0}^{\tau=1}-\frac{1}{c}\int_{0}^{1}
d\tau~w_{b}v^{a}\nabla_{a}v^{b},\label{dsda}\eeq where we have
used that the Lagrangian is given by
$L=(g_{ab}v^{a}v^{b})^{1/2}=c$ along the geodesic path $\gamma$.
Without loss of generality, $w^{a}$ can be chosen orthogonal to
$v^{a}$ and vanishes at end-points to yield the boundary
conditions (\ref{bcwa}). Exploiting the boundary condition
(\ref{bcwa}), the first term in (\ref{dsda}) vanishes and the
least action principle yields the geodesic equation \beq
v^{a}\nabla_{a}v^{b}=0.\label{geod}\eeq

If we have breaks $0=\tau_{0}<\cdots<\tau_{k+1}=1$, and the
restriction of $\gamma$ to each set $[\tau_{i-1},\tau_{i}]$ is
smooth, then the path $\gamma$ is piecewise smooth.  However
$v^{a}$ will generally have a discontinuity at each break
$\tau_{i}$ $(1 \leq i \leq k)$.  This discontinuity is measured by
\beq \Delta v^{a}(\tau_{i})=v^{a}(\tau_{i}^{+})
-v^{a}(\tau_{i}^{-}),\eeq where the first term derives from the
restrictions $\gamma|[\tau_{i},\tau_{i+1}]$ and the second from
$\gamma|[\tau_{i-1},\tau_{i}]$.  If $\gamma$ and $v^{a}\in
T_{\gamma}$ have the breaks $\tau_{1}<\cdots<\tau_{k}$, we have
together with the conditions (\ref{bcwa}) \beq
\sum_{i=0}^{k}v^{a}w_{a}\vert_{\tau_{i}}^{\tau_{i+1}}
=-\sum_{i=1}^{k}w^{a}(\tau_{i})\Delta v^{a}(\tau_{i}),\eeq to
yield \beq S^{(1)}=-\frac{1}{c}\sum_{i=1}^{k}w^{a}(\tau_{i})\Delta
v^{a}(\tau_{i})-\frac{1}{c}\int_{0}^{1}d\tau~w_{b}v^{a}\nabla_{a}v^{b}.
\label{ds0} \eeq Here a path $\gamma\in \Omega(M;p,q)$ is a
critical point of $S$ if and only if the differential
$dS_{\gamma}: T_{\gamma}\Omega(M;p,q)\rightarrow {\mathbf R}$ is
zero, namely along the geodesic
$dS_{\gamma}(w)=\frac{d}{d\alpha}S(\bar{\gamma}(\alpha))=0$ for
every $w\in T_{\gamma}\Omega (M;p,q)$ or every variation
$\bar{\gamma}$ of $\gamma$, if and only if
$\gamma:[0,1]\rightarrow M$ is a geodesic from $p$ to $q$ in $M$.
For a given point $p\in M$ and a tangent vector $v^{a}\in T_{p}M$,
there is a unique geodesic $\gamma_{v}:{\mathbf R}\rightarrow M$
through $\gamma_{v}(0)=p$ whose tangent at $p$ is $v$. The
exponential map $exp_{p}: T_{p}M\rightarrow M$ is defined by
$exp_{p}(v)=\gamma_{v}(1)$.  A point $q\in M$ is conjugate to $p$
if $q$ is a singular value of $exp_{p}: T_{p}M\rightarrow M$.  The
multiplicity of $p$ and $p$ as conjugate points is equal to the
dimension of the null space of $d(exp_{p})_{v}:
T_{v}(T_{p}M)\rightarrow T_{q}M$.

In the stationary phase approximation where $|w_{1}^{a}|$ and
$|w_{2}^{a}|$ are infinitesimally small for $w^{a}_{1},
w^{a}_{2}\in T_{\gamma}\Omega(M;p,q)$, we find the second
variation around the geodesic $\gamma$ \bea S^{(2)}
&=&\frac{\partial^{2}S}{\pa
\alpha_{1}\pa\alpha_{2}}\vert_{\alpha_{1,2}=0}
=-\frac{1}{c}\sum_{i=1}^{k}w_{2}^{b}\na_{b}(w_{1a}(\tau_{i})\Delta
v^{a}(\tau_{i}))\nn\\
&&+\frac{1}{c}\int_{0}^{1}d\tau~g_{ab}w_{1}^{a}\Lambda^{b}_{~c}w_{2}^{c}
\label{dsds} \eea where the Sturm-Liouville operator is given by
\beq\Lambda^{a}_{~b}=-\delta^{a}_{~b}v^{c}\nabla_{c}(v^{d}\nabla_{d})
-R_{cbd}^{~~~a}v^{c}v^{d}.\label{sturm} \eeq If $\gamma$ and
$v^{a}\in T_{\gamma}$ have no breaks, the second variation
$S^{(2)}$ in (\ref{dsds}) vanishes for all $w^{a}_{1}\in
T_{\gamma}\Omega (M;p,q)$ if and only if \beq v^{b}\nabla_{b}
(v^{c}\nabla_{c}w_{2}^{a})+R_{bcd}^{~~~a}v^{b}v^{d}w_{2}^{c}=0.
\label{geodev} \eeq  We call then the $w_{2}^{a}$ a Jacobi field
along $\gamma$.

A point $q$ is conjugate to $p$ along a geodesic $\gamma$ if and
only if there is a non-zero Jacobi field $J$ along $\gamma$ such
that $J(0)=J(1)=0$. Also if $p$ and $q$ are not conjugate along a
geodesic $\gamma$, then a Jacobi field $J$ along $\gamma$ is
determined by its values at $p$ and $q$. A vector field $J\in
T_{\gamma}\Omega (M;p,q)$ is the null space of $S^{(2)}$ if and
only if $J$ is a Jacobi field.  Thus $S^{(2)}$ is degenerate if
and only if $p$ and $q$ are conjugate along $\gamma$. The nullity
of $S^{(2)}$ is equal to the multiplicity (or the dimension of the
space of all Jacobi fields) of $p$ and $q$ as conjugate points. In
the above geodesic deviation equation (\ref{geodev}) we have a
Jacobi field $w_{2}^{a}$ along $\gamma$. A variation
$\bar{\gamma}$ through geodesics $\gamma$ produces a Jacobi field
along $\gamma$ and conversely every Jacobi field is obtained by a
variation of $\gamma$ through geodesics. Define an inner product
on $\Omega (M;p,q)$ by for each $\gamma\in \Omega(M;p,q)$ and
$w_{1}^{a}, w_{2}^{a}\in T_{\gamma} \Omega (M;p,q)$, \beq
(w_{1},w_{2})=\int_{0}^{1}d\tau~g_{ab}w_{1}^{a}w_{2}^{b}. \eeq
With this inner product $T_{\gamma}\Omega(M;p,q)$ will then be a
Hilbert space and $\Omega(M;p,q)$ a Hilbert manifold. The critical
point $\gamma\in\Omega(M;p,q)$ of $S$ is a geodesic from $p$ to
$q$. Its index $ind~(\gamma)$ is defined by the number of points
$\gamma(\tau)$, with $0<\tau<1$, such that $\gamma(\tau)$ is
conjugate to $\gamma (0)$ along $\gamma$, where each conjugate
point is counted with its multiplicity. A geodesic segment
$\gamma:[0,1]\rightarrow M$ contains only finitely many points
which are conjugate to $\gamma(0)$ along $\gamma$, and the
multiplicity of each conjugate point is less than $dim~M=n$. The
space $\Omega(M;p,q)$ of paths from point $p$ to $q$ in $M$, with
the inner product $(~,~)$ is a Hilbert manifold.

\section{Moduli space ${\cal M}(\gamma_{1},\gamma_{2})$}
\setcounter{equation}{0}
\renewcommand{\theequation}{\arabic{section}.\arabic{equation}}

By definition the gradient of the action functional $S:
\Omega(M;p,q)\rightarrow {\mathbf R}$, a vector field $\na S$ on
$\Omega(M;p,q)$ is given by for each $w^{a}\in
T_{\gamma}\Omega(M;p,q)$ \beq dS(w^{a})=w^{a}\nabla_{a}S=\frac{\pa
S}{\pa \alpha} \eeq  which is equivalent to the first variation
(\ref{dsda}). We thus find $dS(w^{a})=0$ for all $w^{a}$ if and
only if $\gamma$ is geodesic. If the path $\gamma$ is smooth and
$\gamma$ and $v^{a}\in T_{\gamma}$ have no breaks, then we obtain
\beq
dS(w^{a})=-\frac{1}{c}\int_{0}^{1}d\tau~w_{a}v^{b}\nabla_{b}v^{a},
\label{ds1} \eeq to yield \beq
\na_{a}S=-\frac{1}{c}\int_{0}^{1}d\tau~v^{b}\nabla_{b}v_{a}.
\label{grads} \eeq

Exploiting (\ref{grads}), we introduce the vector field
$u^{a}=(\pa/\pa \beta)^{a}$ associated with the gradient flow
$\int_{0}^{1}d\tau~\frac{\pa}{\pa
\beta}\bar{\gamma}(\beta,\tau)=-\na S(\bar{\gamma}(\beta))$, where
the trajectory $\bar{\gamma}(\beta)\in \Omega(M;p,q)$ is
identified with the map $\bar{\gamma}: {\mathbf R}\times [0,1]
\rightarrow M$ given by
$\bar{\gamma}(\beta)(\tau)=\bar{\gamma}(\beta,\tau)$ satisfying
\beq u^{a}=\frac{1}{c}v^{b}\nabla_{b}v^{a}. \label{baralpha} \eeq
If for each $\beta\in {\mathbf R}$, $\bar{\gamma}(\beta,\tau)$ is
geodesic, then $\frac{\pa\bar{\gamma}}{\pa \beta}|_{\tau=0}$ is a
Jacobi field and $\bar{\gamma}(0,\tau)=\gamma(\tau)$ a geodesic
path joining $p$ and $q$ in $M$.  If $\bar{\gamma}(\beta,\tau)$
satisfies the asymptotic boundary conditions \beq
\lim_{\beta\rightarrow
-\infty}\bar{\gamma}(\beta,\tau)=\gamma_{1}(\tau),~~~
\lim_{\beta\rightarrow\infty}\bar{\gamma}(\beta,\tau)=\gamma_{2}(\tau)\label{asymbc}\eeq
and $\lim_{\beta\rightarrow \pm\infty}\frac{\pa\bar{\gamma}}{\pa
\beta}(\beta,\tau)=0$ then $\bar{\gamma}(\beta)$ is a trajectory
in $\Omega(M;p,q)$ of the gradient flow joining the geodesic
$\gamma_{1}(\tau)$ and $\gamma_{2}(\tau)$.

There is a natural finite dimensional approximation of the full
path space $\Omega (M;p,q)$, namely for
$0<a_{1}<\cdots<a_{i}<\cdots<a_{\infty}$, let
$\Omega^{a_{i}}=S^{-1}[0,a_{i}]$, then $\Omega^{a_{\infty}}=\Omega
(M;p,q)$.  We choose a subdivision
$0=\tau_{0}<\tau_{1}<\cdots<\tau_{\lambda}=1$ of the unit interval
$[0,1]$.  Let
$\Omega(\tau_{0},\cdots,\tau_{\lambda})=\{\gamma\in\Omega(M;p,q)~|~\gamma(0)=p,
\gamma(1)=q, \gamma|[\tau_{i-1},\tau_{i}]: {\rm
geodesic~for~each}~i=1,\cdots,\lambda\}$, then in fact
$\Omega(\tau_{0},\cdots,\tau_{\lambda})^{a_{i}}=\Omega
(\tau_{0},\cdots,\tau_{\lambda})\cap\Omega^{a_{i}}$ is a finite
dimensional space and $\Omega (M;p,q)$ has a homotopy type of a
countable CW-complex~\cite{milnor63,dold72} which contains one
cell of dimension $ind~(\gamma)$ for each geodesic $\gamma$
(critical point of $S$) in $\Omega (M;p,q)$. Suppose $\gamma_{1}$
and $\gamma_{2}\in \Omega (M;p,q)$ are critical points of the
action functional $S$ with index $k$ and $k-1$, respectively.  By
Sard's theorem, for $p\in M$, almost all $q\in M$ are not
conjugate to $p$ along any geodesic.

We denote by ${\cal M}(M;p,q)$ the space of bounded flow energy
solutions of (\ref{baralpha}), namely \beq {\cal M}(M;p,q)=\{
\bar{\gamma}: {\mathbf R}\times [0,1] \rightarrow M~|~
u^{a}-\frac{1}{c}v^{b}\nabla_{b}v^{a}=0\} \eeq associated with the
bounded flow energy \beq
\Phi(\bar{\gamma})=\int_{-\infty}^{\infty}\int_{0}^{1}d\beta~d\tau~
\left(|u^{a}|^{2}+\frac{1}{c^{2}}|v^{b}\nabla_{b}v^{a}|^{2}\right).
\label{flowe} \eeq The space ${\cal M}(M;p,q)$ may not be compact
in the topology of uniform convergence with all derivative. Assume
the flow is of Morse-Smale type~\cite{palais63}; for every pair
$(\gamma_{1},\gamma_{2})$ of critical points the unstable
submanifold $W^{u}(\gamma_{1})$ and the stable submanifold
$W^{s}(\gamma_{2})$ intersect transversely. \bea &&{\cal
M}(\gamma_{1},\gamma_{2})=\{\bar{\gamma}: {\mathbf R}\times [0,1]
\rightarrow M~|~
u^{a}-\frac{1}{c}v^{b}\nabla_{b}v^{a}=0,\nn\\
&&~~~~~~~~~~~~\lim_{\beta\rightarrow
-\infty}\bar{\gamma}(\beta,\tau)=\gamma_{1}(\tau),
\lim_{\beta\rightarrow\infty}\bar{\gamma}(\beta,\tau)=\gamma_{2}(\tau)\}\nn\\
&&~~~~~~~~~~~~~=W^{u}(\gamma_{1})\cap
W^{s}(\gamma_{2}),\label{mgaga} \eea and the dimension of ${\cal
M}(\gamma_{1},\gamma_{2})$ is given by $dim~{\cal
M}(\gamma_{1},\gamma_{2})=ind~\gamma_{1}-ind~\gamma_{2}$. Moreover
if $dim~{\cal M}(\gamma_{1},\gamma_{2})=1$, then the manifold of
unparametrized trajectories from $\gamma_{1}$ to $\gamma_{2}$,
$\hat{\cal M}(\gamma_{1},\gamma_{2})={\cal
M}(\gamma_{1},\gamma_{2})/{\mathbf R} $ has dimension $0$ and is
compact and orientable.  Here note that \bea
&&\int_{-\infty}^{\infty}\int_{0}^{1} d\beta~d\tau~
|u^{a}-\frac{1}{c}v^{b}\nabla_{b}v^{a}|^{2}\nn\\
&&~~~~~=\Phi(\bar{\gamma})+2S(\gamma_{2})-2S(\gamma_{1}), \eea
where we have used (\ref{grads}). If $\bar{\gamma}\in {\cal
M}(\gamma_{1},\gamma_{2})$, then
$\Phi(\bar{\gamma})=2S(\gamma_{1})-2S(\gamma_{2})$.  Thus ${\cal
M}(\gamma_{1},\gamma_{2})$ is the set of absolute minima of the
function $\Phi$ subject to the asymptotic boundary conditions
(\ref{asymbc}).

For each pair $(\gamma_{1},\gamma_{2})$ of critical points of the
action functional $S: \Omega (M;p,q)\rightarrow {\mathbf R}$, we
have the space of trajectories of the gradient flow of Morse-Smale
type connecting $\gamma_{1}$ and $\gamma_{2}$, namely
(\ref{mgaga}). Every trajectory $\bar{\gamma}: {\mathbf R}\times
[0,1] \rightarrow M$ has a one-dimensional family of
reparametrization $\bar{\gamma}(\lambda+\beta,\tau)$, $\lambda\in
{\mathbf R}$. Denote the space of unparametrized trajectories from
$\gamma_{1}$ to $\gamma_{2}$ by $\hat{\cal
M}(\gamma_{1},\gamma_{2})$.  For a trajectory $\bar{\gamma}:
{\mathbf R}\times [0,1]\rightarrow M$ joining $\gamma_{1}$ and
$\gamma_{2}$, we have the asymptotic boundary conditions
(\ref{asymbc}) and the end point conditions
$\bar{\gamma}(\beta,0)=p$, $\bar{\gamma}(\beta,1)=q$ and the
trajectory $\bar{\gamma}(\beta,\tau)$ satisfies (\ref{baralpha}).

For each nonnegative integer $k$, let $C_{k}$ be the free abelian
group generated by the set of all critical points $\gamma$ with
index $k$ of the action functional $S: \Omega(M; p, q)\ra {\mathbf
R}$.  If $\gamma_{1}\in C_{k}$ and $\gamma_{2}\in C_{k-1}$, then
$\hat{\cal M}(\gamma_{1},\gamma_{2})$ is a zero-dimensional
compact oriented manifold.  Let $n(\gamma_{1},\gamma_{2})$ be the
number of points of $\hat{\cal M}(\gamma_{1},\gamma_{2})$ counted
with the sign of point.  We define, as usual, the boundary
homomorphism \beq \pa_{k}: C_{k}\ra C_{k-1}\eeq by \beq
\pa_{k}(\gamma_{1})=\sum_{\gamma_{2}\in
C_{k-1}}n(\gamma_{1},\gamma_{2})\gamma_{2},\eeq then the
composition of consecutive homomorphism is zero, namely
$\pa_{k-1}\circ\pa_{k}=0$ for all $k$~\cite{biran04} and the
homology group of $\Omega(M;p,q)$ is \beq
H_{k}(\Omega(M;p,q),{\mathbf Z})=\frac{ker~ \pa_{k}:C_{k}\ra
C_{k-1}}{im~ \pa_{k}:C_{k+1}\ra C_{k}}.\eeq For homology theory,
see Refs.~\cite{greenberg81,dold72}.

We denote by $M(S)$ the space of smooth functions $\bar{\gamma}:
{\mathbf R}\times [0,1]\ra M$ which satisfy (\ref{baralpha}) and
have finite flow energy $\Phi (\bar{\gamma})$ in (\ref{flowe}).
Then ${\cal M}(\gamma_{1},\gamma_{2})\subset M(S)$ is the set of
absolute minima of the energy functional $\Phi (\bar{\gamma})$
subject to the asymptotic boundary conditions (\ref{asymbc}). We
consider a vector field $F: M(S)\ra TM(S)$ on $M(S)$ given by for
$\bar{\gamma}\in M(S)$ \beq
F^{a}(\bar{\gamma})=u^{a}-\frac{1}{c}v^{b}\nabla_{b}v^{a}. \eeq
Then ${\cal M}(\gamma_{1},\gamma_{2})\subset F^{-1}(0)$. Moreover
if $\bar{\gamma}\in {\cal M}(\gamma_{1},\gamma_{2})$, then the
projection to the fiber of the differential of $F$ at
$\bar{\gamma}$, $dF_{\bar{\gamma}}: T_{\bar{\gamma}}M(S)\ra
T_{\bar{\gamma}}M(S)$ is given by along the geodesic
$dF_{\bar{\gamma}}(w^{a})=\frac{\pa}{\pa\alpha}F(\bar{\bar{\gamma}})$
where $\bar{\bar{\gamma}}: (-\epsilon,\epsilon)\times{\mathbf
R}\times[0,1] \ra M$ defined by
$\bar{\bar{\gamma}}(0,\beta,\tau)=\bar{\gamma}(\beta,\tau)$. The
map $\bar{\bar{\gamma}}(\alpha,\beta,\tau)$ satisfies
$\bar{\bar{\gamma}}(0,\beta,0)=p$,
$\bar{\bar{\gamma}}(0,\beta,1)=q$ and the asymptotic boundary
conditions $\lim_{\beta\rightarrow
-\infty}\bar{\bar{\gamma}}(0,\beta,\tau)=\gamma_{1}(\tau)$ and $
\lim_{\beta\rightarrow\infty}\bar{\bar{\gamma}}(0,\beta,\tau)=\gamma_{2}(\tau)$.
Moreover along the geodesic we find \beq \frac{\pa F^{a}}{\pa
\alpha}=u^{b}\na_{b}w^{a}+\frac{1}{c}(\Lambda w)^{a}\eeq where
$\Lambda$ is the Sturm-Liouville operator in (\ref{sturm}).  Here
we have used the commutator relations (\ref{poundvw}) and
\beq\pounds_{u}w^{a}=u^{b}\na_{b}w^{a} -w^{b}\na_{b}u^{a}=0.\eeq
For $w_{1}^{a},w_{2}^{a}\in T_{\bar{\gamma}}M(S)$, we have
$L^{2}$-inner product on $T_{\bar{\gamma}}M(S)$, \bea
(w_{2},dF_{\bar{\gamma}}(w_{1}))&=&\int_{-\infty}^{\infty}\int_{0}^{1}d\beta~d\tau~g_{ab}\nn\\
&&\cdot\bar{w}_{2}^{a}\left(\bar{u}^{c}\na_{c}\bar{w}_{1}^{b}+\frac{1}{c}(\Lambda
\bar{w}_{1})^{b}\right) \eea where $\Lambda$ is the
Sturm-Liouville operator in (\ref{sturm}), and
$\bar{w}_{i}^{a}=\bar{\gamma}^{*}w_{i}^{a}$ $(i=1,2)$ and
$\bar{u}^{a}=\bar{\gamma}^{*}u^{a}$.

\section{Example on the spheres}
\setcounter{equation}{0}
\renewcommand{\theequation}{\arabic{section}.\arabic{equation}}

We consider a particle of mass $m$ on the $n$-sphere $S^{n}$ in a
conservative physical system where the total energy $E$ is
constant and the potential energy $V$ depends only on the radial
distance $r$ from the center of the sphere $S^{n}$, such as a
particle on an $S^{2}$ in an attractive gravitational potential of
$V(r)$.  In this case both the metrics $\eta_{ab}$ and $g_{ab}$ in
(\ref{metric}) are just metrics for $S^{n}$. Suppose that two
points $p$ and $q$ in $S^{n}$ are neither identical nor antipodal.
Then there are countably many geodesics $\gamma_{0}$,
$\gamma_{1}$, $\cdots$ from $p$ to $q$ in $S^{n}$. Here let
$\gamma_{0}$ be the shortest great circle arc $pq$ from $p$ to
$q$, let $\gamma_{1}$ be the long circle arc $p(-q)(-p)q$, let
$\gamma_{2}$ be the arc $pq(-p)(-q)pq$, and so on.  The set
$C(S)=\{\gamma_{0}, \gamma_{1},\cdots\}$ is the critical points of
the action functional $S: \Omega(S^{n};p,q)\ra {\mathbf R}$. The
subscript $k$ of $\gamma_{k}$ is the number of times that $p$ or
$(-p)$ lies in the interior of $\gamma_{k}$. Each of the points
$p$ or $(-p)$ in the interior of $\gamma_{k}$ is conjugate to $p$
with multiplicity $(n-1)$.  The path space $\Omega (S^{n};p,q)$
has the homotopy type of a CW-complex
structure~\cite{milnor63,dold72} with one cell each in the
dimension $0$, $n-1$, $2(n-1)$, $\cdots$.  Using the CW-complex
structure and the trajectory of gradient flow of $S$, we may
compute the homology groups of $\Omega (S^{n};p,q)$.  For $n\ge
3$, since all boundary map of the complex
$C_{*}(\Omega(S^{n};p,q))\stackrel{\pa}{\ra}
C_{*-1}(\Omega(S^{n};p,q))$ is zero, the homology group
$H_{k(n-1)}(\Omega (S^{n};p,q))={\mathbf Z}$, $k=0,1,2,\cdots$.
For $n=1$, the path space $\Omega (S^{1};p,q)$ has countably many
connected components. Each component of them is contractible and
has minimum action functional at the unique geodesic in its
component. The path space $\Omega (S^{1};p,q)$ has the homotopy
type of a CW-complex structure with countably many zero-cells.
Thus the homology group of $\Omega (S^{1};p,q)$ is the group of
countable direct sum of ${\mathbf Z}$ at zero dimension;
$H_{0}(\Omega (S^{1};p,q))=\oplus {\mathbf Z}$.  For $n=2$ the
path space $\Omega (S^{2};p,q)$ has the homotopy type of a
CW-complex structure with one cell in each dimension. The singular
complex of $\Omega (S^{2};p,q)$ is \beq \cdots\ra C_{k}=\langle
\gamma_{k}\rangle={\mathbf Z}\stackrel{\pa}{\ra}C_{k-1}=\langle
\gamma_{k-1}\rangle= {\mathbf Z}\ra\cdots. \eeq The space $\Omega
(S^{2};p,q)$ has only one connected component since $S^{2}$ is
simply connected.  (In fact, so is $\Omega (S^{n};p,q)$ if $n>2$.)
The moduli space $\hat{\cal M}(\gamma_{k},\gamma_{k-1})$ has two
distinct unparameterized trajectories with opposite orientations.
Thus $\pa_{k}=0$ for all $k\ge 0$ and the homology groups of
$\Omega (S^{2};p,q)$ are \beq H_{k}(\Omega (S^{2};p,q))={\mathbf
Z},~~~{\rm for~all~}k\ge 0. \eeq

\begin{table}[h]
\caption{Homologies of $S^{n}$ and $\Omega (S^{n};p,q)$}
\begin{center}
\begin{tabular}{|c|ccccc||c|ccccccc|}
\hline  & & &dim & & & & & & &dim & & & \\
\cline{2-6} \cline{8-14}
space &0 &1 &2 &3 &4  & space &0 &1 &2 &3 &4 &5 &6\\
\hline $S^{0}$ &${\mathbf Z}\oplus{\mathbf Z}$ &0 &0 &0 &0
&$\Omega(S^{1})$ &$\oplus{\mathbf Z}$
&0 &0 &0 &0 &0 &0\\
$S^{1}$ &${\mathbf Z}$ &${\mathbf Z}$ &0 &0 &0 &$\Omega(S^{2})$
&${\mathbf Z}$
&${\mathbf Z}$ &${\mathbf Z}$ &${\mathbf Z}$ &${\mathbf Z}$ &${\mathbf Z}$ &${\mathbf Z}$\\
$S^{2}$ &${\mathbf Z}$ &0 &${\mathbf Z}$ &0 &0 &$\Omega(S^{3})$
&${\mathbf Z}$
&0 &${\mathbf Z}$ &0 &${\mathbf Z}$ &0 &${\mathbf Z}$\\
$S^{3}$ &${\mathbf Z}$ &0 &0 &${\mathbf Z}$ &0 &$\Omega(S^{4})$
&${\mathbf Z}$
&0 &0 &${\mathbf Z}$ &0 &0 &${\mathbf Z}$\\
$S^{4}$ &${\mathbf Z}$ &0 &0 &0 &${\mathbf Z}$ &$\Omega(S^{5})$
&${\mathbf Z}$
&0 &0 &0 &${\mathbf Z}$ &0 &0\\
\hline
\end{tabular}
\end{center}
\end{table}

The real projective space ${\mathbf R}{\mathbf P}^{\infty}$ of
dimension $\infty$ has also a CW-complex structure with one cell
in each dimension.  However the homology group $H_{k}({\mathbf
R}{\mathbf P}^{\infty})$ of ${\mathbf R}{\mathbf P}^{\infty}$ is
different with the one of $\Omega (S^{2};p,q)$.  Thus $\Omega
(S^{2};p,q)$ and ${\mathbf R}{\mathbf P}^{\infty}$ are not
homotopically equivalent.  The homology of real projective
space~\cite{milnor74} is \beq H_{k}({\mathbf R}{\mathbf
P}^{\infty})=\left\{
\begin{array}{ll}
{\mathbf Z} &{\rm if}~k=0\\
{\mathbf Z}_{2} &{\rm if}~k>0~{\rm is~odd}\\
0 &{\rm if}~k>0~{\rm is~even.}
\end{array}\right.
\eeq

\section{Conclusions}
\setcounter{equation}{0}
\renewcommand{\theequation}{\arabic{section}.\arabic{equation}}

In conclusion, the action functional for a point particle has been
introduced in a conservative physical system with constant total
energy to formulate the geodesic equation together with breaks on
the path. By taking the second variation of the action functional,
the geodesic deviation equation has been derived and discussed in
terms of the Jacobi fields on the curved manifold.

Defining the gradient of the action functional, the space of
bounded flow energy solutions has been investigated to construct
the moduli space associated with the critical points of the action
functional and the asymptotic boundary conditions. The boundary
homomorphism has been also introduced to define the homology group
of the path space.  We have considered the particle motion on the
$n$-sphere $S^{n}$ in the conservative physical system to discuss
the moduli space of the path space and the corresponding homology
groups.

Applying the Morse theoretic approach developed for the point
particle to the string theory, one could consider the gradient of
the string action functional and the moduli space associated with
the critical strings of the string action functional. It would be
also desirable if the homology group of the stringy tube space can
be studied in the framework of the Morse theory. These works are
in progress and will be reported elsewhere.

\acknowledgments The work of YSC was supported by the Korea
Research Council of Fundamental Science and Technology (KRCF),
Grant No. C-RESEARCH-2006-11-NIMS, and the work of STH was
supported by the Korea Research Foundation (MOEHRD), Grant No.
KRF-2006-331-C00071, and by the Korea Research Council of
Fundamental Science and Technology (KRCF), Grant No.
C-RESEARCH-2006-11-NIMS.

\end{document}